\documentclass[a4paper,fleqn]{cas-sc}

\usepackage[authoryear,longnamesfirst]{natbib}
\usepackage{xcolor}
\usepackage{tabularx} 
\usepackage{ragged2e} 
\usepackage{array}    
\usepackage{times}

\def\tsc#1{\csdef{#1}{\textsc{\lowercase{#1}}\xspace}}
\tsc{WGM}
\tsc{QE}
\tsc{EP}
\tsc{PMS}
\tsc{BEC}
\tsc{DE}

\begin{document}
\let\WriteBookmarks\relax
\def\floatpagepagefraction{1}
\def\textpagefraction{.001}

\shorttitle{Data Architectures for AI-Ready Interoperable Public Transportation Ecosystems}

\shortauthors{}

\title [mode = title]{Data Architectures for AI-Ready Interoperable Public Transportation Ecosystems}                      



%
\author[1]{Diego Da Silva}



\ead{dluiz.silva@utoronto.ca}



\affiliation[1]{organization={Department of Civil \& Mineral Engineering, University of Toronto},
    city={Toronto},
    country={Canada}}

\author[2]{Raphael Y. de Camargo}
\ead{raphael.camargo@ufabc.edu.br}


\author[3]{Mayuri A. Morais}

\affiliation[2]{organization={Center for Mathematics, Computing and Cognition, Universidade Federal do ABC},
    city={Santo André},
    state={SP},
    country={Brazil}}

\affiliation[3]{organization={FGV Cidades, Fundação Getulio Vargas},
    city={São Paulo},
    country={Brazil}}

\author[1]{Amer S. Shalaby}
\ead{amer.shalaby@utoronto.ca}



\nonumnote{}

\begin{abstract}Public transportation (PT) agencies generate vast amounts of heterogeneous data from automatic fare collection (AFC), automatic passenger counting (APC), vehicle location (AVL/CAD), schedule and real-time feeds (GTFS/GTFS-RT), and proprietary platforms. These datasets offer unprecedented opportunities for data-driven planning, operations, and passenger services, but their potential is constrained by fragmentation, inconsistent update frequencies, and the lack of reproducible, interoperable pipelines. While contemporary data platform patterns and architectural styles from enterprise computing address analogous challenges in other sectors, their adaptation to the PT domain remains mostly underexplored. Transit systems present unique conditions, including the convergence of Information Technology (IT) and Operational Technology (OT), long asset lifecycles, rigorous security requirements, multi-agency coordination requirements, and the need to operate on live systems that preclude controlled experimentation.
  
This paper analyzes the domain-specific data architecture challenges confronting transit agencies and proposes the \emph{Transit Data Integration Spectrum}, which classifies transit data-architecture positions along a continuum from \emph{Segregated} (institution-centric) through \emph{Coordinated} (agency-coordinated, domain-based) to \emph{Federated} (cross-agency, governed). For each position, we characterize ten facets: dominant ownership pattern, deployment and access model, agency context, operational posture, cybersecurity posture, interoperability scope, AI-readiness and analytics capability, governance model, role of standards, and organizational model. We further examine the role of open data standards as architectural contracts that facilitate progression along the spectrum, as well as the legacy constraints that shape realistic migration pathways. The spectrum provides transit agencies and researchers with a basis for assessing current data-architecture maturity and for identifying feasible paths toward more interoperable, AI-ready data ecosystems. 
\end{abstract}







\maketitle

\section{Introduction}
\label{sec:intro}

Public transportation (PT) is currently at a pivotal stage in its digital evolution, reflecting the global technological revolution driven by advancements in Artificial Intelligence (AI) \citep{jevinger2024artificial-982,Petrozziello}. Recent examples show PT systems using Large Language Models (LLMs), computer vision, and predictive modeling for passenger assistance, crowd monitoring, and fare-evasion detection \citep{UITP2025AI}. This transformation is enabled by data that is collected, integrated, and analyzed on an unprecedented scale, supported by sensing technologies and the Internet of Things (IoT) \citep{SAKI2026100242}. The pressure for architectural change in PT systems extends beyond new computing technologies. PT increasingly operates within a mobility ecosystem converging around technology-driven platforms, including Transportation Network Companies (TNCs) \citep{Diao2021}, micromobility providers \citep{Oeschger2020Micromobility,Cui2024SharedMicromobility}, and Mobility as a Service (MaaS) platforms \citep{Jittrapirom2017MaaS}. TNCs have intensified competition for passengers, using highly optimized, data-centric infrastructures to attract riders, particularly in contexts where fixed-route services are perceived as inflexible or insufficient \citep{Diao2021,Hanig2025}. Finally, in metropolitan regions, PT frequently involves multiple agencies that depend on real-time data exchange and cross-agency coordination to improve operational efficiency and passenger experience \citep{Juster2015RegionalRealtimeTransit,Gkiotsalitis2023TransferSync}. Nonetheless, for Transit Agencies (TAs), the primary challenge lies in adopting architectural design choices that enable these capabilities \citep{AReferenceArchitecture2024}.

Historically, PT architecture has evolved alongside the broader modernization of Intelligent Transportation Systems (ITS), which accelerated after the rapid motorization that began in the 1940s \citep{HistoryITS2021}. However, the emphasis has been on Systems Engineering (SE) rather than Enterprise Architecture (EA), with static reference architectures that generally follow layered, monolithic designs, such as the Architecture Reference for Cooperative and Intelligent Transportation (ARC-IT) \citep{NASEM2011TEAP,ARCIT}. This approach offers both stability and opportunities for standardization, facilitating systematic procurement processes, enabling modular software performance assessment, and preventing redundant component deployment across different organizational scales. However, as we will discuss in this paper, this static, often monolithic design can limit effective data integration and AI-readiness \citep{HAEFNER2023122878,welch2019big,kai21:bigdatatransit,NAP26674}. The path forward requires elevating architecture to a strategic priority that recognizes the unique institutional, operational, and collaborative nature of PT \citep{Hemily2015TransitITS}, and that empowers architectural and data teams as strategic contributors rather than custodians of back-office systems.

\begin{figure}[htp!]
  \centering
  \includegraphics[width=0.7\textwidth]{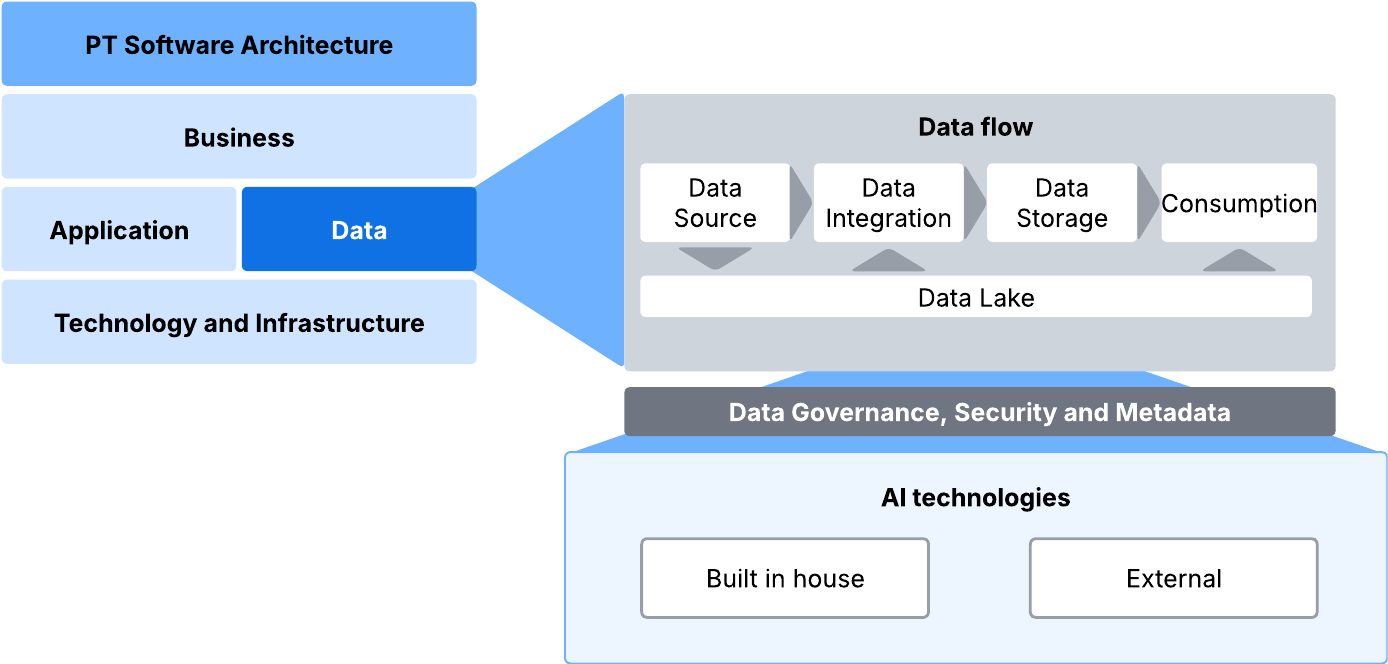}
  \caption{Example of an enterprise architecture}\label{fig:diag7}
\end{figure}

In many industry sectors, the subsystems, applications, and functional modules that support service management are organized within an Enterprise Architecture (EA) framework, which provides organization-wide architectural governance and strategic alignment linking institutional objectives, capabilities, and systems \citep{NASEM2011TEAP,OpenGroup_TOGAF_Leaders}, as illustrated in Figure \ref{fig:diag7}. EA defines the strategic intent and governance that guides organizational transformation. The data architecture indicates how data is ingested, stored, governed, exchanged, and served across operational and analytical contexts. It can be based on different architecture styles, which denote recurring structural patterns such as layered, event-driven, or domain-oriented designs \citep{Kruchten,richards2020fundamentals} for structuring systems components and data pipelines. Finally, different data platform types can be used for data storage and processing, such as warehouses, lakes, and lakehouses \citep{azzabi24:datalakesurvey,Armbrust2021Lakehouse}. Despite industry adoption, PT has been slow to implement EA practices, mainly due to limited resources and a lack of strategic backing \citep{NASEM2011TEAP}, with only 2 of 17 surveyed agencies using them. 

Two key concepts we use in this work are interoperability and AI-readiness. We define interoperability as the capacity of systems within a transit data architecture to exchange and reuse data across organizational and technical boundaries through shared semantics, standardized interfaces, and explicit trust and governance arrangements \citep{EIF2017,CEN_Transmodel,NASEM2020DataSharingGuidance}. We define AI-readiness as the architectural capacity to support reliable access to historical data for training and streaming data for inference, with integration across operational domains and capabilities for repeatable deployment, monitoring, and auditability of AI workflows \citep{Huyen,Huyen2,Kreuzberger2023MLOps,Paleyes2022}. 

\begin{figure}[htp!]
  \centering
  \includegraphics[width=1\textwidth]{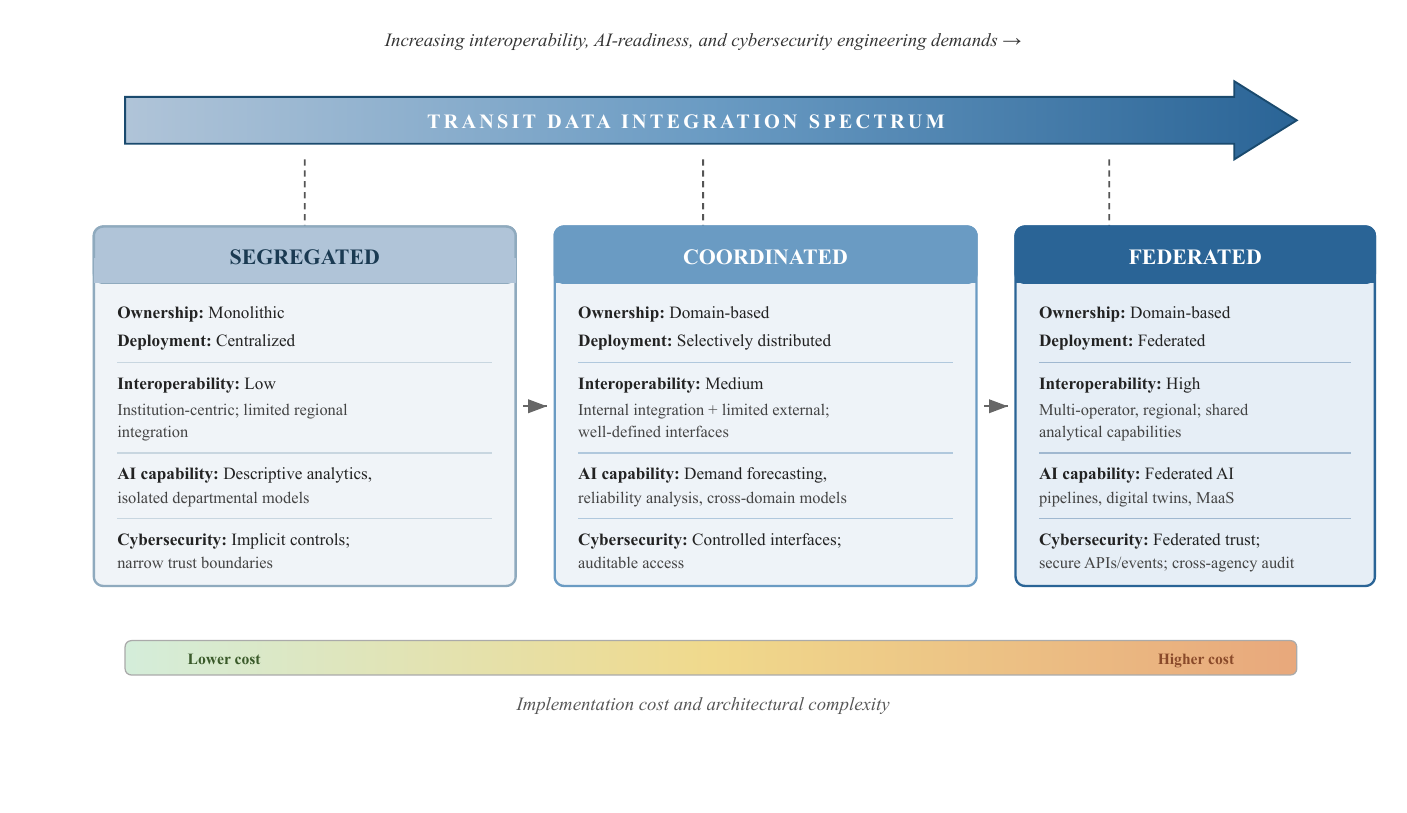}
  \caption{The Transit Data Integration Spectrum, with architectural choices along a continuum from Segregated (institution-centric, siloed-domains) to Federated (multiple agencies or operators), with increasing interoperability, AI-readiness, and implementation complexity.}\label{fig:diag3}
\end{figure}

In this work, we present a \emph{Transit Data Integration Spectrum} of architectural choices available to TAs along a continuum of increasing interoperability and AI-readiness (Figure~\ref{fig:diag3}). The spectrum was derived through a structured synthesis, by comparing recurrent data-architecture patterns from enterprise computing and software engineering \citep{Kruchten, richards2020fundamentals} with transit-specific constraints documented in the ITS, transit planning, and data standards literature \citep{NASEM2011TEAP, AReferenceArchitecture2024, Babar2019,Guerreiro2016,Zhu2019,Amini2017}. We identified three architectural positions (segregated, coordinated, and federated) and characterized each position across ten practitioner-oriented facets. 

Within this analytical model, the \emph{Segregated} end represents the most common architecture: siloed, monolithic, vendor-centered systems in which data is typically optimized for a single subsystem. This configuration is robust and simple to deploy, but it offers limited interoperability and extensibility. \emph{Coordinated} configurations introduce domain-level organization within a single agency, with clearly defined boundaries and interfaces, shared semantics, and explicit governance. This organization improves reliability and maintainability, enabling cross-domain application deployment and internal coordination. At the \emph{Federated} end, multi-agency architectures provide interoperability across operators and agencies through shared semantics, identifiers, security, and governance.

This paper brings the following contributions:
\begin{itemize}
    \item Proposes the Transit Data Integration Spectrum as a framework for understanding architectural choices in PT data ecosystems, with three levels: Segregated, Coordinated, and Federated.
    \item Characterizes each position along the spectrum across ten practitioner-oriented facets: dominant ownership pattern, deployment and access model, agency context, operational posture, cybersecurity posture, interoperability scope, AI-readiness and analytics capability, governance model, role of standards, and organizational model.
    \item Positions open standards (GTFS, NeTEx, SIRI, TIDES) as semantic and trust contracts, not only external publishing formats.
    \item Synthesizes transit-specific constraints, including OT/IT convergence, cybersecurity, legacy constraints, and the role of standards, that influence system migration toward greater interoperability and AI-readiness.
\end{itemize}

\section{Related Work}
\label{sec:related-work}

The relevant literature covers topics related to transit data models, transit data sharing, and ITS architectures. 

\subsection{Transit Data Models and Standards}

TRANSMODEL is the European Reference Data Model for Public Transport~\citep{CEN_Transmodel}. It provides an abstract, implementation-independent data model that defines entities, relationships, attributes, and semantics across the major functional domains of PT, including network topology, scheduling, fare management, operations monitoring, passenger information, and personnel disposition. It ensures that concepts such as journey, service pattern, or stop place carry consistent meaning across systems and organizations. Some specifications derived from TRANSMODEL include NeTEx for planned data, SIRI for real-time data, and, more recently, OpRa for observed operational data~\citep{Data4PT_Models}. These standards constitute a mature model for transit data exchange in Europe and are increasingly referenced internationally.

The General Transit Feed Specification (GTFS), developed through a collaboration between TriMet and Google, provides a lightweight, practically motivated specification that has been adopted by many TAs around the globe to provide transit schedules and real-time information~\citep {Antrim2013GTFS}. However, GTFS focuses on passenger-facing information rather than internal operational management and is normally used mostly as an external publishing format. 

More recently, the Transit Integrated Data Exchange Specification (TIDES) has been proposed as an open, community-governed specification for historical transit operations data~\citep{TIDES2024}. Rather than defining a single flat feed, TIDES organizes event tables such as vehicle locations, passenger events, and fare transactions together with summary tables such as stop visits, trips performed, and station activities. It also includes explicit metadata, validation rules, and links observed operations to the scheduled service. TIDES targets the operational data that agencies need for internal use and can serve as a standardized internal and cross-agency data exchange format.

Other works address the conditions under which transit data can be shared and reused. For instance, some works examine institutional objectives, licensing, benefits, risks, and governance arrangements for sharing transit data \citep{NASEM2015OpenData,NASEM2020DataSharingGuidance}. Another line of work addresses data quality and conformance in widely used exchange standards, showing that GTFS feeds often contain schema and content errors and therefore require validation and best-practice guidance to remain dependable inputs for analytics and passenger information systems \citep{Devunuri2024GTFSErrors,gtfs_schedule_best_practices,gtfs_realtime_best_practices}. A third line of research addresses privacy-preserving transit data release, especially for farecard and trajectory-like records, and the use of specific mechanisms, such as minimization, de-identification, and formal privacy \citep{Pelletier2011SmartCard,Dempsey2008SmartCardPrivacy,Chen2012DifferentialPrivacyTransit,Garroussi2025MaaSPrivacy}. 

These approaches address data modeling, data exchange, sharing governance, data quality, and privacy. Our work complements theirs by addressing data architectures and how these standards and data-sharing practices can be applied at all levels of the spectrum. In particular, we argue that data standards should be understood not only as formats for external exchange but also as architectural contracts that organize internal data flows, service interfaces, and organizational coordination. 

\subsection{Architectures for Intelligent Transportation Systems}

The Architecture Reference for Cooperative and Intelligent Transportation (ARC-IT)~\citep{ARCIT} is a comprehensive systems engineering reference architecture for ITS in the United States. ARC-IT defines physical and logical subsystem boundaries, information flows, and interface standards across the ITS domain, including transit, traffic management, emergency services, and traveler information. It provides a standardized vocabulary for ITS system integration and procurement, and embeds security directly into the ITS architecture. However, ARC-IT reflects a static, n-layered view of system design, specifying what subsystems exist and how they communicate. 

\cite{AReferenceArchitecture2024} proposed a reference architecture for data-driven Intelligent Public Transportation Systems (IPTS), from the experience of deploying a large-scale system with Hitachi Rail. Their architecture identifies domain-specific requirements for IPTS and proposes a layered design that addresses data ingestion, processing, and analytics within a single-system deployment. This work shows that generic smart-city architectures are insufficient for PT and that domain-specific architectural guidance is needed. Our paper differs in two respects. 

At the federated end of the spectrum, data architecture requirements are increasingly shaped by the demands of Mobility as a Service (MaaS). \cite{Jittrapirom2017MaaS} provided an early characterization of MaaS, identifying core attributes including integration of transport modes, personalization, tariff options, and platform-based service delivery. They show that MaaS platforms depend on real-time, multi-operator data access, federated identity and payment systems, and governance frameworks that span organizational boundaries. A recent work shows that cybersecurity has been under-treated, reinforcing that regional interoperability creates new trust and security obligations rather than only new data-sharing opportunities \citep{Peralta2024MaaSCyber}. 

Our contribution complements the existing literature by providing a spectrum of data architectures for PT with different levels of interoperability and AI-readiness, extending beyond single-system design to address multi-operator interoperability and regional coordination.

\section{The Need for a Domain-Specific Transit Data Architecture}
\label{sec:domain-specific-arch}

PT operates under conditions that differ fundamentally from those of most other data-intensive sectors \citep{CUTA1993Handbook, Hemily2015TransitITS}. The differences stem not only from institutional and regulatory environments but also from the coexistence of Operational Technology (OT) and Information Technology (IT). On the one hand, OT includes vehicle control systems, signaling equipment, fare validators, power systems, and wayside sensors \citep{ARCIT,NIST80082r3}. These components are purpose-built, often vendor-specific, expected to function continuously for many years, sometimes decades \citep{KHATTAK199639}, and embedded in physical infrastructure, directly affecting service delivery and passenger safety. On the other hand, IT typically supports business processes, analytics, planning, and user-facing services. Devices and platforms in this category are usually standardized, commercially available, and designed for relatively short replacement cycles. Software environments evolve quickly, updates are frequent, and failures are often recoverable without immediate safety or service consequences. 

From a data architecture perspective, IT and OT are tasked with creating and maintaining the TA's primary data sources that inform most strategic decision-making \citep{NAP13917,Ge2021TransitDataSources}. OT generates and consumes data under tighter reliability and timing constraints, such as signaling and fare validation, which require high-throughput ingestion and immediate processing. Some of this data is further transformed into formats useful for IT, such as historical records of vehicle locations, passenger counts, and validated fare transactions, which can be used for planning, AI model training, and performance evaluation \citep{jevinger2024artificial-982,SAKI2026100242}. Other data is produced and used entirely within IT, such as scheduling, planning, customer relationship management, and financial data. Because OT and IT impose different constraints on latency, reliability, and data structure, they typically rely on distinct staff teams and architectural solutions, with limited integration between them. The resulting separation is not a design choice but an inherited consequence of these competing requirements. These differences make it challenging to adopt off-the-shelf data architecture patterns developed from other industries \citep{AReferenceArchitecture2024}. Moreover, the system should provide secure OT-IT boundaries, since IT platforms increasingly expose data for analytics, APIs, and passenger-facing services, while OT assets directly affect service delivery and passenger safety \citep{NIST80082r3,APTA2022Cybersecurity, APTA2023OTCMF}.

TAs face an additional constraint that system changes must take effect on live, safety-critical systems that serve passengers in real time, with limited ability to deploy parallel versions, perform controlled experiments, or rollback quickly \citep{NIST80082r3,APTA2022Cybersecurity}. Moreover, OT assets in transit can have life cycles measured in decades, and replacement cycles are influenced by capital planning and procurement timelines \citep{KHATTAK199639,NIST80082r3}. Any architectural change must therefore coexist with legacy systems for extended periods, requiring incremental, modular introduction that does not destabilize ongoing operations \citep{NASEM2020DataSharingGuidance,NAP26674}.

AI models are initially trained on historical data, but after deployment, they use streaming data to make real-time predictions \citep{Huyen,Sculley2015}, requiring architectures that support both batch processing and streaming ingestion \citep{Armbrust2021Lakehouse,Kreuzberger2023MLOps}. At the same time, AI models can consume data from multiple domains, such as fare, vehicle location, traffic, and external weather data \citep{Sculley2015,fta2022emerging}. These multiple domain, streaming requirements make authenticated APIs and event channels, data-integrity checks, and auditability prerequisites for trustworthy interoperability \citep{NIST800207}.

Regarding data sharing, PT agencies may need to collaborate with multiple operators, regional authorities, and external partners to enable integrated services, regional planning, and policy evaluation \citep{NASEM2020DataSharingGuidance,NASEM2015OpenData,Mahajan2022DataToThePeople}. A domain-specific data architecture must therefore balance local operational control with the ability to exchange data across institutional boundaries, a distinction from industries where data is treated primarily as a competitive asset. Also, data sharing must consider privacy constraints, as fare collection records and passenger identities can expose sensitive personal, socio-economic, and mobility pattern information \citep{Pelletier2011SmartCard,Dempsey2008SmartCardPrivacy,Garroussi2025MaaSPrivacy}. Collectively, these challenges compose a complex and ever-evolving puzzle, where multiple valid solutions exist but poorly designed ones risk producing what software architecture literature calls a \textit{spaghetti architecture}, namely a tightly entangled system that is difficult to maintain, scale, or extend.

\section{Data Architecture Strategies for PT}
\label{sec:data-arch-strategies}

Data architectures for TAs must provide operational reliability, security, and institutional control in environments that include multiple OT and IT systems from heterogeneous vendors. These constraints lead to transit systems with siloed architectures, limited interoperability, and low AI-readiness. In this section, we discuss transit data architectures with different levels of interoperability and AI-readiness.

\begin{figure}[htp!]
  \centering
  \includegraphics[width=.6\textwidth]{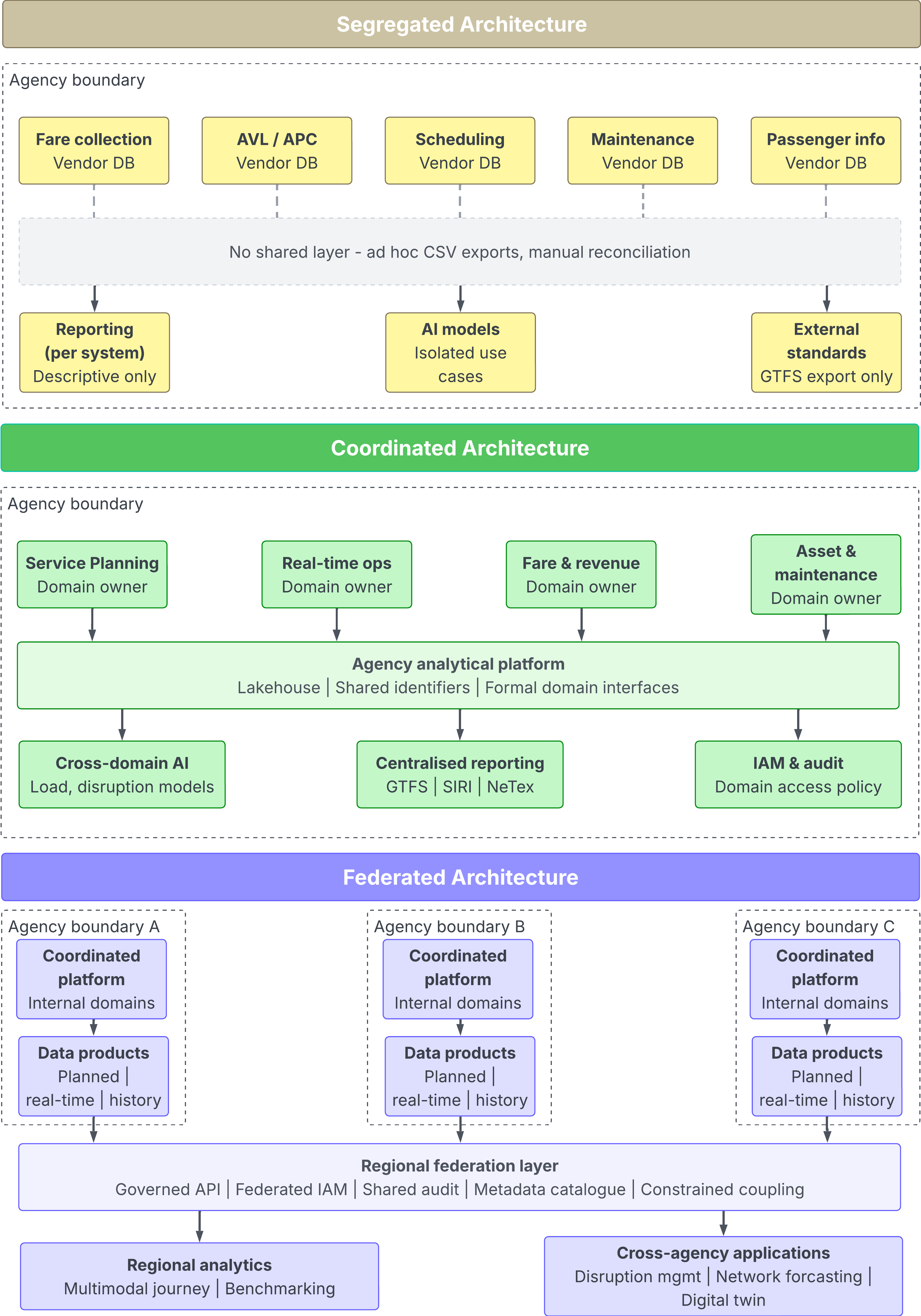}
  \caption{Data architecture strategies for PT.}\label{fig:diag_archi}
\end{figure}

TAs can analyze data architecture decisions along three dimensions that reflect the desired ownership granularity, deployment and access model, and governance and interoperability scope, as shown in Table~\ref{table-data-arch}. Ownership granularity indicates whether data is managed as a monolithic institutional asset, divided into major domains, or organized into narrow service-specific stores \citep{Goedegebuure2024DataMesh}. The deployment and access model indicates whether data is stored on a single platform, selectively distributed within a single agency, or federated among multiple platforms \citep{richards2020fundamentals,azzabi24:datalakesurvey,Armbrust2021Lakehouse}. Finally, the governance and interoperability scope concerns whether standards, identifiers, access rules, and change management are embedded within individual systems, coordinated within a single agency, or formalized through multiple agencies and operators \citep{NASEM2020DataSharingGuidance,NIST800207}. 

\begin{table}[ht]
\centering
\scriptsize
\caption{Three Dimensions of Data Organization and Coordination in Transit Systems}
\renewcommand{\arraystretch}{1.25}
\begin{tabularx}{\textwidth}{|p{3.2cm}|X|X|}
\hline
\textbf{Dimension} & \textbf{Categories} & \textbf{What the dimension influences} \\
\hline
\textbf{Data ownership granularity} &
\textbf{Monolithic}: all data is managed within a tightly coupled platform. \newline
\textbf{Domain-based}: each major transit domain owns data and interfaces. \newline
\textbf{Per-service}: narrow services own highly specific data stores. &
Data ownership, system coupling, accountability, maintainability, and data sharing. \\
\hline
\textbf{Deployment and access model} &
\textbf{Centralized}: a single primary database or platform handles most functions. \newline
\textbf{Selectively distributed}: high-volume or latency-sensitive domains are separated, but remain under a single agency control. \newline
\textbf{Federated}: multiple autonomous platforms or agencies expose access through shared contracts, catalogs, APIs, queries, or events. &
Scalability, locality, failure isolation, trust boundaries, segmentation, monitoring, and cross-boundary access. \\
\hline
\textbf{Governance and interoperability scope} &
\textbf{System-centric}: semantics, access rules, and change management are embedded within individual systems or vendors. \newline
\textbf{Agency-coordinated}: identifiers, metadata, and interfaces are aligned within a single agency. \newline
\textbf{Cross-agency federated}: semantics, trust, audit, and versioning are coordinated among operators or authorities. &
How data sharing, metadata, identity, access control, audit, supplier trust, and incident coordination are managed within and across agencies. \\
\hline
\end{tabularx}
\label{table-data-arch}
\end{table}

These dimensions are complementary and admit many combinations, such as domain-based, centralized, and system-centric, or per-service, centralized, and agency-coordinated. For clarity, however, we synthesize the combinations most relevant to PT into three positions, shown in Figure~\ref{fig:diag_archi}. At the \emph{Segregated} end, we typically see small to medium-sized agencies with stable operations and constrained budgets, where data analytics is performed mostly within individual departments. At the \emph{Coordinated} level, domains are separated but aligned through shared semantics and consistent internal interfaces, enabling easier deployment of multiple domain applications within an agency. At the \emph{Federated} end, applications involving multiple agencies, such as regional planning, shared situational awareness, coordinated passenger information, benchmarking, and decision support, become possible through shared data semantics and regional governance. The levels are therefore distinguished mostly by the scope of desired interoperability and governance, while specific technologies provide the means to achieve these objectives.

In practice, discussion of data architecture often centers on platform patterns such as warehouses, data lakes, and lakehouses, and on architectural styles such as layered, event-driven, or domain-oriented. Conventional data warehouses are valuable for structured reporting and historical analysis \citep{richards2020fundamentals}, while data lakes expand support for semi-structured and high-volume sources \citep{azzabi24:datalakesurvey}, and lakehouses combine lake flexibility with warehouse-like governance \citep{Armbrust2021Lakehouse}. Regarding architectural styles, layered architectures support separation of concerns through clearly differentiated functional responsibilities \citep{Kruchten,richards2020fundamentals}, event-driven architectures support decoupling, real-time monitoring, and operational responsiveness \citep{Laigner}, and domain-oriented architectures emphasize ownership around stable business domains and stronger coordination of data as a product \citep{Goedegebuure2024DataMesh}. These architectural styles can be applied at most levels of the spectrum and used simultaneously within the same system for different types of data and use cases.

Another important point is cybersecurity, which should be a cross-cutting concern. For instance, ARC-IT embeds security into the ITS architecture \citep{ARCIT}, while NIST OT guidance and transit-sector practice frame defense-in-depth, segmentation, asset inventory, governance, procurement and lifecycle risk management as architectural necessities for systems that must remain safe and continuously available \citep{NIST80082r3,NISTCSF2024,APTA2022Cybersecurity, IEC62443, CENELECTS50701}.

\begin{table}[ht]
\centering
\scriptsize
\caption{Transit Data Integration Spectrum: Practitioner-Oriented Characterization of Public Transportation Data Ecosystems}
\renewcommand{\arraystretch}{1.30}
\setlength{\tabcolsep}{4pt}
\begin{tabularx}{\textwidth}{|>{\raggedright\arraybackslash}p{2.7cm}|>{\raggedright\arraybackslash}X|>{\raggedright\arraybackslash}X|>{\raggedright\arraybackslash}X|}
\hline
\textbf{Facet} & \textbf{Segregated} & \textbf{Coordinated} & \textbf{Federated} \\
\hline
\textbf{Dominant ownership pattern} &
System- or vendor-centered. Data is typically owned by individual applications such as AFC, AVL/CAD, scheduling, or asset systems. &
Domain-based within one agency. Major domains own data products and interfaces, but treat data as shared institutional assets. &
Agency-autonomous, regionally federated. Each agency retains ownership of its local systems while exposing governed domain data products under shared regional contracts. \\
\hline
\textbf{Deployment and access model} &
Centralized system databases or multiple siloed databases. Integration is usually performed through custom ETL, manual reconciliation, or vendor-specific connectors. &
Integrated storage platform with well-defined data schemas for each domain, typically using a centralized lakehouse-like platform and, where needed, distributed storage for high-volume or latency-sensitive domains. &
Multiple autonomous agency platforms connected through governed APIs, event exchange, shared catalogs, federated query, and selective replication where needed. \\
\hline
\textbf{Agency context} &
Small to medium agencies, prioritizing operational continuity, vendor support, and low change risk. &
Medium to large agencies, with improved analytics capabilities and interoperability while protecting core operational systems. &
Large metropolitan regions with multiple agencies and service integration. \\
\hline
\textbf{Operational posture} &
Each operational area operates mostly in isolation, with clear boundaries, resulting in reliable operations. However, interactions involving different domains are brittle and ad hoc. &
Mission-critical operations remain protected behind controlled interfaces, while applications involving multiple domains become more reliable and easier to deploy. &
Mission-critical operations remain local, while data sharing, planning, and coordination involving multiple agencies become possible.\\
\hline
\textbf{Cybersecurity posture} &
Controls are local and often implicit. Security relies heavily on vendor defaults and isolation, with uneven visibility, patching, and logging across systems. &
Security is engineered through data classification, access management, segmented OT-IT interfaces, auditable access, centralized monitoring, and explicit data stewardship. &
Security requires multi-party trust governance, federated identity and access management, secure APIs and event mediation, shared audit, supplier risk controls, coordinated incident response, and resilience against cascading failures. \\
\hline
\textbf{Interoperability scope} &
Low. Mostly institution-centric. Cross-domain and external exchange depend on bespoke interfaces, bilateral agreements, or periodic publications. &
Medium. Reusable internal contracts and selected external exchanges are established, but most coordination remains centered on a single agency. &
High. Cross-agency integration is possible through shared semantics, access policies, and versioning rules. \\
\hline
\textbf{AI-readiness and analytics capability} &
Low AI-readiness. Historical data is fragmented, cross-domain integration is limited, and AI pipelines are difficult to maintain. Analytics remain mostly descriptive and isolated within individual departments or systems. &
Moderate AI-readiness. Cross-domain access and model development governance within a single agency become feasible for applications such as reliability analysis, demand forecasting, planning, and real-time operational control and monitoring. &
High AI-readiness for regional use cases. Federated multi-operator access, shared semantics, and reusable interfaces make cross-agency analytics, forecasting, passenger information, and decision-support pipelines more feasible. \\
\hline
\textbf{Governance model} &
Implicit and centralized within systems, vendors, or small IT groups. Metadata, lineage, and data quality practices are limited or inconsistent. &
Agency-wide governance with domain stewardship, metadata management, access rules, and more explicit quality controls. &
Federated governance across agencies, including semantics, access, audit, and standards. \\
\hline
\textbf{Role of standards} &
Primarily external reporting or compliance formats. Standards may be published outwardly, but seldom structure internal data flows or interface security practices. &
Standards begin to function as internal interface and trust contracts for key domains, improving consistency, authentication, versioning, and audit capabilities. &
Standards help provide the semantic and trust foundations for federation across agencies, enabling interoperable data products, more consistent regional analytics, and a governed, secure exchange. \\
\hline
\textbf{Organizational model} &
Siloed teams aligned to systems, departments, or vendors. Coordination costs are high, and knowledge remains siloed. &
Hybrid model combining a central platform with per-domain governance, with higher upfront investment but better long-term support for cross-domain interoperability.&
Local agency teams remain in place, complemented by shared regional governance and interoperability functions rather than a fully centralized regional IT organization. \\
\hline
\end{tabularx}
\label{tab:spectrum}
\end{table}

Table~\ref{tab:spectrum} characterizes the Transit Data Integration Spectrum across ten practitioner-relevant facets that agencies can use to select an appropriate architectural target. Within the proposed analytical model, the spectrum can be applied through four diagnostic questions. First, where is authoritative ownership for core transit data actually defined: within individual systems, within agency domains, or through cross-agency contracts? Second, how are new analytical or integration use cases delivered: through bespoke interfaces, reusable internal contracts, or reusable regional data products? Third, who controls data governance, including semantics, access, audit, and standards? Fourth, how are trust boundaries, access decisions, monitoring, and incident coordination handled: as local system concerns, agency-wide controls, or shared regional capabilities? Agencies whose main priority is enabling applications across internal domains through governed interfaces will often align with the coordinated position. At this level, regional integration is also possible, but each use case still requires custom negotiation and specific data pipelines. In larger metropolitan areas, where regional integration is a strategic priority, the federated approach may be appropriate because it enables the easier integration of multiple agencies and operators through reusable trust and interoperability arrangements.

\subsection{Segregated Architecture: Siloed, Vendor-Centered, and Institution-Centric Systems}

The segregated strategy is the most common and, in many legacy environments, the most practical configuration. Data and applications are organized primarily around individual systems, such as fare collection, scheduling, vehicle location, passenger information, maintenance, and customer management. Vendor platforms frequently include their own data stores, interface logic, and reporting tools, making local operation dependable but leaving little incentive to treat data as a shared asset. The architecture is typically centralized at each system level, even when the agency operates several such systems in parallel. This position persists not only because of technical debt, but because it satisfies important transit objectives. Local accountability is clear, trust boundaries are relatively narrow, certification and change management are simpler, and operational teams can prioritize service continuity over extensibility. For safety- and revenue-critical functions, such conservatism is often rational.

From a cybersecurity perspective, segregation can reduce exposure by reducing the number of external interfaces and keeping many OT systems locally isolated. However, it often leaves agencies with uneven logging, patching, remote-access control, and incident handling across vendor systems \citep{APTA2022Cybersecurity,NIST80082r3}. Security may appear stronger because boundaries are narrow, but in practice, it is frequently contingent on inherited vendor defaults and ad hoc compensating controls.

Beyond technical considerations, the choice of data architecture strategy has profound implications for team organization, governance, and institutional capacity within PT agencies. Architectural decisions shape not only how systems interact but also how people collaborate, how responsibilities are distributed, and how effectively agencies can operate at local and regional scales. Under the segregated data architecture strategy, technical and organizational fragmentation tend to reinforce one another \citep{NASEM2020DataSharingGuidance}. Monolithic and centralized architectures are typically accompanied by siloed team structures, where data management, IT, and operational technology functions are separated by domain, vendor, or departmental boundaries. These teams often focus on maintaining specific systems rather than managing shared data assets or cross-domain workflows. As a result, coordination costs are high, institutional knowledge is fragmented, and innovation depends heavily on individual projects rather than collective capability. In this context, AI initiatives are frequently treated as specialized or experimental efforts, isolated from core operational teams and difficult to scale beyond their original scope.

This position represents a recurring pattern in transit practice, with the coexistence of AVL/APC/AFC, scheduling, passenger information, and customer management systems procured separately and managed through vendor-specific interfaces, periodic extracts, or manual reconciliation. Analytics, therefore, remain mostly descriptive, governance often remains implicit within systems and vendors, standards function largely as external reporting formats, and organizational structures remain siloed rather than oriented toward shared data stewardship \citep{fta2022emerging,NASEM2020DataSharingGuidance}.

For agencies at this position, the first step should be to take low-risk actions to reduce vendor lock-in. In particular, they can establish global identifiers for routes, stops, trips, vehicles, and assets, improve data exportability, build asset and interface inventories, better document interfaces, require minimum logging and export capabilities, tighten remote and administrative access controls, and require open or at least mappable formats alongside cybersecurity obligations in procurements \citep{gtfs_schedule_best_practices,gtfs_realtime_best_practices,NASEM2020DataSharingGuidance,NIST80082r3,APTA2023OTCMF}. These steps do not eliminate segregation, but they create the preconditions for more deliberate transition, as we discuss in Section~\ref{sec:legacy}.

\subsection{Coordinated Architecture: Domain-Based Coordination Within a Single Agency}

The coordinated strategy characterizes many agencies actively modernizing their data environments. Its defining feature is the domain-based organization within a single agency. Core data is grouped around relatively stable business and operational domains, such as service planning and scheduling, real-time operations, fare and revenue, customer information, asset and maintenance, and enterprise support. Unlike segregated architectures, the agency also defines clear data ownership, shared identifiers, and formal interfaces between domains \citep{Goedegebuure2024DataMesh}. This organization enables the deployment of cross-domain analytical tools and AI applications.

This structure can be implemented through a centralized or selectively distributed platform. The simplest solution is to introduce an agency-wide analytical platform, such as a lakehouse, to decouple reporting, model development, and historical analysis from the operational area. High-volume or latency-sensitive domains such as AVL, APC, or AFC may be supported through dedicated pipelines or infrastructure, while lower-volume functions remain on the shared platforms. What matters is that technical boundaries are aligned with domain responsibilities and that internal reuse becomes systematic rather than ad hoc.

As agencies move toward coordinated architectures, organizational structures begin to evolve alongside technical decomposition. The introduction of domain-based systems and distributed components often necessitates closer collaboration between data engineering, IT, and operational teams. While these teams may still be formally distinct, shared interfaces, domain contracts, and analytical pipelines create incentives for coordination. Coordinated architectures, therefore, encourage hybrid organizational models in which centralized governance coexists with domain-level ownership \citep{Goedegebuure2024DataMesh,fta2022emerging}. However, without deliberate management strategies, this structure can also introduce ambiguity in roles and accountability, as technical decoupling outpaces organizational realignment.

This position is especially attractive in PT because it improves maintainability without unnecessarily risking core operations. Stable OT systems can remain protected behind controlled interfaces while new pipelines, metadata practices, and model workflows are introduced alongside them. Security becomes more explicit through data classification, domain identity and access management, policy enforcement at APIs and event brokers, auditable access controls, centralized logging and monitoring for shared analytical platforms, segmented OT-IT interfaces, and clearer stewardship responsibilities \citep{NIST800207,OWASP2023API,NIST80082r3}. Interoperability also improves because standards such as GTFS, GTFS-RT, SIRI, NeTEx, or TIDES can begin to function as internal interface contracts. Finally, cross-domain AI becomes feasible, since reliability analysis, passenger load prediction, real-time disruption response, and operational coordination can rely on data from multiple domains \citep{welch2019big,kai21:bigdatatransit}.

One example of coordinated architecture is provided by \cite{AReferenceArchitecture2024}, which was derived from a large-scale intelligent PT deployment in an Italian city and emphasizes integrated ingestion, processing, and analytics within one urban system. In that system, domains are technically integrated, and analytics are shared, but control remains centralized in a single system.

A critical distinction follows from this definition. A region composed of several agencies with strong coordinated architectures is \emph{not} yet federated if each new cross-agency use case still requires bespoke negotiation, custom ETLs, or one-off loading into a regional hub. Coordinated maturity addresses internal fragmentation and prepares an agency for operation in a federated architecture, which is described in the next subsection.

\subsection{Federated Architecture: Cross-Agency Federation with Constrained Coupling}

The federated data architecture enables collaboration between multiple agencies and operators in the same region. The idea is that each agency retains local ownership of its systems and data products while also participating in a regional federation, enabling the development and deployment of applications that involve multiple agencies and operators. This is achieved through common semantics, identifier mapping, metadata discovery, access control, audit requirements, and versioning practices, which facilitate inter-agency deployments. The defining characteristic of the federated architecture is not the specific mechanism, but the fact that cross-agency coordination becomes repeatable and governed. 

In this paper, the recommended federated model for PT is constrained coupling. Federation is most valuable for regional analytics, passenger information, performance management, service planning, benchmarking, and selected coordination functions such as disruption visibility or integrated journey support. Constrained coupling is also a security decision: by keeping signaling, dispatching, fare validation, and other mission-critical functions local, agencies reduce the scope of impact and limit cascading failures when shared layers are degraded or attacked \citep{NIST80082r3,APTA2016OCSZ}. This distinction is required to make regional integration compatible with transit requirements for reliability, security, and institutional autonomy.

Two possible architectural styles are domain-based federation and per-service federation. In the former, each agency exposes a limited number of domain-specific data products, such as planned service, real-time operations, historical operations, disruptions, and fare data managed under explicit privacy constraints \citep{Pelletier2011SmartCard,Dempsey2008SmartCardPrivacy,Chen2012DifferentialPrivacyTransit,Garroussi2025MaaSPrivacy}. In the latter, each service, such as scheduling, AVL, fare collection, or passenger information, is federated separately across agencies, enabling more deeply integrated operations that can improve efficiency and passenger experience. However, this approach requires significant reorganization within each agency and increases coupling across operations, expanding the scope of failure and the risk of cascading disruptions across agencies. Consequently, this paper treats domain-based federation as the recommended default for PT under the constraints discussed here, while recognizing that per-service federation remains possible for selected use cases. This recommendation reflects the view that domain-based federation better preserves resilience, institutional autonomy, and incremental adoption while still delivering many of the benefits of regional integration.

Federated architectures can support richer analytical and AI use cases by making multi-operator data accessible under agreed-upon rules. Applications such as regional origin-destination estimation, cross-operator diagnostics, multimodal passenger information, coordinated disruption management, and network-level forecasting become more feasible \citep{Juster2015RegionalRealtimeTransit,Gkiotsalitis2023TransferSync,NASEM2020DataSharingGuidance}. Shared data environments can also support more advanced observability and simulation workflows, including transit digital twin applications, when agencies adopt sufficiently governed and interoperable data pipelines \citep{DeBenedictis2024DigitalTwinIPTS}. At the same time, this position is the most institutionally demanding. Its costs arise less from technology than from the need for shared governance, funding, security engineering, legal agreements, sustained stewardship, supplier management, and cross-agency incident coordination. The use of domain-based federation reduces the scope of cross-agency coordination, since local systems are affected only where they interface with the federated layer, allowing incremental adoption of the federated architecture.

Cybersecurity becomes even more relevant on federated architectures, since the attack surface area increases with multiple APIs and data sharing among operators. In this scenario, federated identity and access management, shared logging and audit schemas, secure API and event mediation, supplier risk controls, and incident escalation procedures must become part of the architecture from its inception and procurement process \citep{Peralta2024MaaSCyber,NIST800207,OWASP2023API}.

One example at this position is the Washington, DC experience with regional real-time transit communications and data sharing \citep{Juster2015RegionalRealtimeTransit}, where shared regional tools improved situational awareness and coordination, yet participating agencies retained local systems and mission-critical operational control, making this a partial federation.

To summarize, the integration spectrum can be seen as a roadmap for organizational integration, enabling agencies to move from fragmented, system-centric operations to cohesive, regionally coordinated mobility management. By consolidating expertise and responsibility across regions, agencies can shift from reactive system maintenance toward proactive system management. The appropriate level for each agency depends on its context, including the region in which it operates, which may range from a small city to a large metropolitan area, as well as its budget, institutional capacity, and strategic priorities. The provided guidelines should help agencies decide which level to target and understand the capabilities and limitations of each level.

\section{Transit System Migration and the Role of Standards}
\label{sec:legacy}

Legacy systems are a defining characteristic of PT agencies worldwide. Many essential transit functions, such as scheduling, fare collection, vehicle tracking, and asset management, rely on systems designed decades ago, based on assumptions that differ significantly from today's. These systems are deeply integrated into agency workflows, certified under strict regulatory standards, and closely linked to operational technology that must function continuously and safely. Additionally, staff members who have managed these systems for many years possess specialized expertise and established workflows that, while valuable, can make them resistant to changes. Consequently, transitioning to coordinated and federated data architectures involves complex technical, organizational, and institutional challenges.

Two characteristics of existing transit systems are particularly consequential for data architecture: vendor lock-in and siloed subsystems. Vendor lock-in occurs when vendors deliver end-to-end solutions using proprietary formats and interfaces. Data produced by these systems may be stored in closed databases, exported only as static reports, or embedded in formats intended for human interpretation rather than machine processing. In some cases, agencies have difficulty accessing their own data, particularly when vendor contracts limit extraction or documentation is incomplete. The second related issue is the presence of siloed subsystems, with scheduling, vehicle location, fare collection, and passenger information systems frequently procured independently, each optimized for its own operational purpose. As a result, similar data elements, such as stop identifiers, route definitions, and service patterns, may be represented differently across systems. Over time, these inconsistencies become institutionalized, reinforcing the siloed practices that characterize the segregated end of the spectrum \citep{Mahajan2022DataToThePeople,NASEM2020DataSharingGuidance,NAP26674, Hemily2015TransitITS}.

Within this context, data specifications and standards can play a role that goes beyond data exporting. Specifications such as GTFS, NeTEx, SIRI, and TIDES provide structured representations of transit data that, when adopted internally, can function as semantic and trust contracts. They can be used to standardize data in multiple subsystems, reduce ambiguity, and establish common semantics. Mapping existing data structures to standardized schemas forces agencies to confront inconsistencies and gaps, revealing where similar concepts are treated differently and where critical attributes are missing or poorly defined \citep{Devunuri2024GTFSErrors,gtfs_schedule_best_practices}. That mapping effort must also be paired with authentication, authorization, auditability, and versioning responsibilities at the interfaces where data is exposed or reused \citep{NIST800207,OWASP2023API}, and, when sharing data between sectors and agencies, with privacy-by-design obligations for fare, identity, and journey data \citep{NASEM2020DataSharingGuidance,Chen2012DifferentialPrivacyTransit,Garroussi2025MaaSPrivacy}. While this effort is demanding and often reveals significant data quality issues that require dedicated remediation, it is what enables agencies to operate at the coordinated and federated position in the transit data integration spectrum.

Adopting standards-based architectures under these constraints requires a phased migration strategy. A possible approach is to run new standards-compliant data pipelines in parallel with existing systems, using intermediate conversion layers, so that agencies can validate outputs against operational ground truth without disrupting service delivery \citep{richards2020fundamentals,NAP26674}. As systems are renewed through procurement cycles, agencies can require native support for open standards, reducing dependence on conversion layers over time \citep{samtrafiken2024standards,NASEM2020DataSharingGuidance}. 

\section{Conclusions}


Transit data architectures in most PT agencies remain siloed, institution-centric, and tightly coupled to proprietary vendor implementations. These characteristics directly limit the adoption of AI capabilities, constrain interoperability, and inhibit the regional coordination that urban mobility increasingly expects. However, transit practitioners frequently lack the guidelines to adopt architectures that enable these emerging applications.

In this paper, we analyzed the unique challenges that distinguish transit data architecture from enterprise data management in other domains, including the coexistence of IT and OT, regulatory constraints on systems, and the institutional complexity of multi-agency environments \citep{AReferenceArchitecture2024,NASEM2011TEAP}. To address these challenges, we proposed the Transit Data Integration Spectrum, which positions architectural strategies along a continuum from Segregated (institution-centric) through Coordinated (agency-coordinated, domain-based) to Federated (cross-agency). Each position was characterized on ten facets: dominant ownership pattern, deployment and access model, agency context, operational posture, cybersecurity posture, interoperability scope, AI-readiness and analytics capability, governance model, role of standards, and organizational model. The analysis provides a structured basis for agencies to assess their current position and identify realistic evolution paths. 

We also identified the role of open data specifications such as GTFS, NeTEx, SIRI, and TIDES as potentially semantic and trust contracts that, when adopted internally, can be used to standardize data semantics across subsystems, improving both interoperability and AI-readiness \citep{Antrim2013GTFS,Data4PT_Models,TIDES2024,NIST800207}. 

This work also has some limitations. The Transit Data Integration Spectrum has not been empirically validated through case studies or structured expert evaluation. Also, fully federated architectures remain a conceptual target and are only partially realized in the literature reviewed \citep{Juster2015RegionalRealtimeTransit}. Finally, the spectrum necessarily abstracts over contextual variation in agency size, institutional capacity, regulatory environment, and regional governance structures, all of which influence how architectural strategies are adopted in practice.

Our next step is to develop a concrete reference architecture that specifies how data flows through ingestion, storage, processing, governance, serving, and consumption stages at the coordinated and federated positions, including identity, trust boundaries, monitoring, supplier governance, and incident coordination requirements. Empirical validation through case studies of agencies at different points along the spectrum could then test the framework's descriptive accuracy and practical utility.

\section*{Acknowledgments}
Any use of generative AI in this manuscript adheres to ethical guidelines for use and acknowledgement of generative AI in academic research. Each author has made a substantial contribution to the work, which has been thoroughly vetted for accuracy, and assumes responsibility for the integrity of their contributions. 



\printcredits

\bibliographystyle{cas-model2-names}
\bibliography{cas-refs}



\end{document}